# Full-Duplex Relay with Jamming Protocol for Improving Physical-Layer Security

Saeedeh Parsaeefard and Tho Le-Ngoc
Electrical and Computer Engineering Department, Mcgill University
saeideh.parsaeifard@mail.mcgill.ca and tho.le-ngoc@mcgill.ca

*Abstract*—This paper proposes a jointly cooperative relay and jamming protocol based on full-duplex (FD) capable relay to increase the source-destination secrecy rate in the presence of different types of eavesdroppers. In this so called *FD-Relay with jamming (FDJ)* protocol, the FD-Relay, first, simultaneously receives data and sends jamming to the eavesdropper, and, then, forwards the data, while the source jams the eavesdropper. Achievable secrecy rates of the proposed FDJ in presence of different eavesdropper types and self-interference (SI) are derived and compared with those of the traditional half-duplex (HD) relay. The adaptive power allocation for secrecy rate maximization in a multi-carrier scenario for both proposed FDJ and HD-Relay is formulated as a non-convex optimization problem and corresponding iterative solution algorithm is developed using the <u>d</u>ifference-of-two-<u>c</u>oncave-functions (DC) programming technique. The simulation results confirm that FDJ offers significant improvements in the secrecy rate over the HD-Relay.

## I. Introduction

To improve the coverage and capacity of next-generation wireless networks, relay has been recognized based on two schemes: *I)* Half-duplex (HD) relaying where relay R receives data from source S, then, R forwards data to the destination (D); *II)* Full-duplex (FD) relaying where R receives and sends simultaneously. The advantage of FD-Relay mode to increase the spectral efficiency has been reported, e.g., [1], where the performance comparison of FD and HD-Relay highly depends on the self-interference (SI) of FD-Relay.

In addition, next-generation wireless networks deal with the vulnerable broadcasting feature of wireless channels against overhearing of eavesdroppers (E), where the secrecy rate is chosen as the figure of merit, and defined as the difference between the achieved S-D rate and eavesdropper overheard rate [2]. Obviously, when S-D channel gain is lower than S-E channel gain, the secrecy rate is zero, indicating unfavorable interference-limited scenario [3]. In this context, to increase the chance of non-zero secrecy rate, applying the HD-Relay to increase S-D desired rate is a natural approach e.g., [4]. Also, it is well studied that the external jamming nodes can be deployed to degrade the overheard rate [5], and also relay and jamming nodes can jointly improve S-D secrecy rate, e.g., [6].

This paper aims to deploy the FD capability of relay to propose the cooperative relay and jamming protocol to increase the S-D secrecy rate, called *FD-Relay with jamming (FDJ)* protocol. The benefits and even disadvantages of FD-capable nodes on secrecy rate have been initiated in [3], [7].

We investigate the potential benefit of FD-Relay in improving S-D secrecy rate where R concurrently increases S-D desired rate and decreases the overheard rate of eavesdropper. In the proposed FDJ, in the first time slot, the relay simultaneously receives from S and sends jamming to E. Then, in the second time slot, R forwards the data to D, while S sends jamming to E, meaning that the FD-Relay offers jointly cooperative transmission and jamming.

The achievable secrecy rate depends on the capability of the eavesdroppers [4], categorized into two types: 1) Naive eavesdropper who just extracts the information from either S or R; and 2) Informed eavesdropper who decodes the information from both S and R. We derive the secrecy rates of FDJ for these two types which allow us to further establish the conditions for non-zero secrecy rates for FDJ, and compare the FDJ and HD-Relay secrecy rates. Our analysis indicates that FDJ can improve the secrecy rate compared to the HD-Relay and the level of SI has a strong influence on the FDJ secrecy rates. We further investigate the FDJ power allocation problems. Based on the <u>d</u>ifference-of-two-<u>c</u>oncave-functions (DC) programming, we develop an efficient iterative algorithm to solve these non-convex optimization problems. Simulation results confirm the analytical results and show that the FDJ can improve the secrecy rate up to twice that offered HD-Relay for small level of SI.

The rest of this paper is organized as follows. Section II describes the system model, followed by Section III where a performance comparison analysis of FDJ and HD-Relay is presented. Section IV contains the power allocation problems and their solutions. Section V provides the simulation results and Section VI concludes the paper.

## II. Network Model and Problem Formulation

Consider a two-hop transmission between a single-antenna S and D via a friendly FD-Relay R in the presence of E shown in Fig. 1. The total bandwidth of $B$ Hz allocated to S-D transmission, is divided into $K$ equal-bandwidth sub-carriers forming $\mathcal{K} = \{1, \cdots, K\}$. Assume that $B/K$ is much less than the channel coherence bandwidth so that the sub-carrier frequency response is flat and represented by the channel power gain $h_{XY}^k$ for sub-carrier $k$ over the X-Y link where $X \in \{S, R\}$ and $Y \in \{R, D, E\}$. Consequently, the overall channel responses over the bandwidth $B$ of X-Y link are represented by $\mathbf{h}_{XY} = [h_{XY}^1, \cdots, h_{XY}^K]$.

The transmission frame consists of two equal time-slots denoted by $t = 1$ and $t = 2$. The channel gains are assumed to be block fading. Also, we assume that $h_{SD}^k \approx 0$ and R uses the decode-and-forward (DF) strategy. In the HD-Relay protocol illustrated in Fig. 1(a), in $t = 1$, the source transmits $x^k[t=1]$ to R over sub-carrier $k$ where $E\{|x^k[t=1]|^2\} = P_S^k$, $E\{.\}$ is the expectation operator, and $P_S^k$ is transmit power of S on sub-carrier $k$. The received signals at R and E from S on sub-carrier $k$ are, respectively, $r_{HD}^k[t=1] = \sqrt{h_{SR}^k} x^k[t=1] + n_R^k$ and $e_{HD}^k[t=1] = \sqrt{h_{SE}^k} x^k[t=1] + n_E^k$ where $n_R^k$ and $n_E^k$ are the white Gaussian noise samples at R and E, respectively for all $k \in \mathcal{K}$. Without loss of generality, we assume the same noise power of $\sigma$ at each receiver in all sub-carriers. In $t = 1$, the signal-to-noise-power ratio (SNR) at R and E are $\gamma_R^k(\text{HD}) = \frac{P_S^k h_{SR}^k}{\sigma}$ and $\gamma_{SE}^k(\text{HD}) = \frac{P_S^k h_{SE}^k}{\sigma}$, respectively, for all $k \in \mathcal{K}$. In $t = 2$, R forwards $f^k[t=2]$ to D, and the received signals at D and E are $y_{HD}^k[t=2] = \sqrt{h_{RD}^k} f^k[t=2] + n_D^k$, $e_{HD}^k[t=2] = \sqrt{h_{RE}^k} f^k[t=2] + n_E^k$, respectively, where $E\{|f^k[t=2]|^2\} = P_R^k$, $P_R^k$ is the transmit power of R on sub-carrier $k$, and $n_D^k$ is the white Gaussian noise samples at D. At D, the SNR is $\gamma_D^k(\text{HD}) = \frac{P_R^k h_{RD}^k}{\sigma}$, and, the SNR for the HD-Relay with DF strategy is $\gamma_{HD}^k = \min\{\gamma_R^k(\text{HD}), \gamma_D^k(\text{HD})\}$, for all $k \in \mathcal{K}$ [8]. In $t = 2$, the overheard SNR of E is $\gamma_{RE}^k(\text{HD}) = \frac{P_R^k h_{RE}^k}{\sigma}$. For the naive eavesdropper with only listening capability without knowledge of relay operation, its effective SNR is $\gamma_E^{k,1} = \max\{\gamma_{SE}^k(\text{HD}), \gamma_{RE}^k(\text{HD})\}$, for all $k \in \mathcal{K}$, and consequently, the S-D secrecy rate in the case of a naive eavesdropper is $\text{SR}^1 = \sum_{k=1}^K \text{SR}^{k,1}$ where $\text{SR}^{k,1} = [\log 2(1 + \gamma_{HD}^k) - \log 2(1 + \gamma_E^{k,1})]^+$, and $[x]^+ = \max\{0, x\}$. However, for an informed eavesdropper with knowledge of relay operation, its effective SNR is $\gamma_E^{k,2} = \frac{P_S^k h_{SE}^k}{\sigma} + \frac{P_R^k h_{RE}^k}{\sigma}$, and consequently, the S-D secrecy rate in the case of an informed eavesdropper is $\text{SR}^2 = \sum_{k=1}^K \text{SR}^{k,2}$ where $\text{SR}^{k,2} = [\log 2(1 + \gamma_{HD}^k) - \log 2(1 + \gamma_E^{k,2})]^+$.

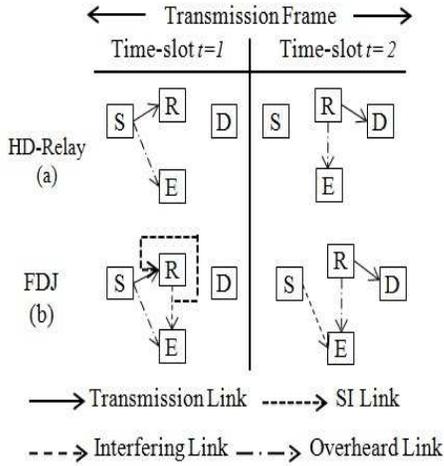

Fig. 1. Operation in one transmit frame (a) HD-Relay and (b) FDJ.

Using FDJ, in $t = 1$, S sends data to R, and, simultaneously, R tries to jam the overhearing of E by sending artificial noise $\tilde{f}^k[t=1]$ where $E\{|\tilde{f}^k[t=1]|^2\} = P_R^k$. Therefore, in time-slot $t = 1$, the received signal at R is $r^k[t=1] = \sqrt{h_{SR}^k} x^k[t=$ $1] + \sqrt{h_{SI}^k} \tilde{f}^k[t=1] + n_R^k$, for all $k \in \mathcal{K}$, and its SINR is $\gamma_R^k(t=1) = \frac{P_S^k(1) h_{SR}^k}{\sigma + P_R^k(1) h_{SI}^k}$ where we use 1 to indicate $t = 1$ for brevity. In $t = 2$, S sends the artificial noise $\tilde{x}^k[t=2]$ to jam E and, simultaneously, R forwards the data to D. The active links in $t = 1$ and $t = 2$ are depicted in Fig. 1 (b). The received signal at D is $y^k[t=2] = \sqrt{h_{RD}^k} f^k[t=2] + n_D^k$, and, consequently, its SINR is $\gamma_D^k(t=2) = \frac{P_R^k(2) h_{RD}^k}{\sigma}$. Here, we use 2 to indicate $t = 2$ for brevity. Consequently, SINR of FDJ for the DF strategy, is $\gamma_{FDJ}^k = \min\{\gamma_R^k(t=1), \gamma_D^k(t=2)\}$. In this protocol, the overhearing signal of E in $t = 1$ and $t = 2$ are $e^k[t=1] = \sqrt{h_{SE}^k} x^k[t=1] + \sqrt{h_{RE}^k} \tilde{f}^k[t=1] + n_E^k$, $e^k[t=2] = \sqrt{h_{SE}^k} \tilde{x}^k[t=2] + \sqrt{h_{RE}^k} f^k[t=2] + n_E^k$, for all $k \in \mathcal{K}$. For the naive eavesdropper, the overheard SINR in $t = 1$ and $t = 2$ are $\gamma_{SEJ}^k(t=1) = \frac{P_S^k(1) h_{SE}^k}{\sigma + P_R^k(1) h_{RE}^k}$ and $\gamma_{REJ}^k(t=2) = \frac{P_R^k(2) h_{RE}^k}{\sigma + P_S^k(2) h_{SE}^k}$. Consequently, we have $\gamma_E^{k,3} = \max\{\gamma_{SEJ}^k(t=1), \gamma_{REJ}^k(t=2)\}$ for all $k \in \mathcal{K}$. In this case, S-D secrecy rate with a naive eavesdropper is $\text{SR}^3 = \sum_{k=1}^K \text{SR}^{k,3}$, where $\text{SR}^{k,3} = [\log 2(1 + \gamma_{FDJ}^k) - \log 2(1 + \gamma_E^{k,3})]^+$. When the eavesdropper is informed, assuming same codebook at S and R, its overheard SINR is $\gamma_E^{k,4} = \frac{P_S^k(1) h_{SE}^k}{\sigma + P_R^k(1) h_{RE}^k} + \frac{P_R^k(2) h_{RE}^k}{\sigma + P_S^k(2) h_{SE}^k}$, and we have $\text{SR}^4 = \sum_{k=1}^K \text{SR}^{k,4}$ where $\text{SR}^{k,4} = [\log 2(1 + \gamma_{FDJ}^k) - \log 2(1 + \gamma_E^{k,4})]^+$, for all $k \in \mathcal{K}$. FDJ is similar to the case with one friendly HD-Relay and one friendly jammer in the network, except that here FD-Relay acts as both friendly relay and jammer with SI.

## III. PERFORMANCE COMPARISONS

In this section, we study the achievable secrecy rates of FDJ, and compare them with HD-Relay. We first focus on the single-carrier transmission scenario where the index $k$ can be dropped for simplicity without confusion. We also assume no power allocation is applied to S and R, i.e., they transmit at their maximum transmit power. Without loss of generality, consider $P_S^{\max} = 1$, and $\alpha = P_R^{\max}/P_S^{\max}$. Let assume that the noise is negligible compared to interference, e.g., $\sigma \ll P_R^{\max} h_{SI}$ and $\sigma \ll P_S^{\max} h_{SE}$. As previously mentioned, one concern for secrecy rate is that it might be zero depending on the channel gains. In the following, we show how FDJ can increase the chance of non-zero secrecy rate compared to the HD-Relay protocol.

*Non-Zero Secrecy Rate Conditions for a Naive Eavesdropper:* For the naive eavesdropper, by considering $\gamma_E^1 < \gamma_{HD}$ and $\gamma_E^3 < \gamma_{FDJ}$, the conditions to achieve non-zero secrecy rate for the FD and HD-Relay protocols (i.e., for $\text{SR}^1$ and $\text{SR}^3 > 0$) can be derived as follows:

C01: For HD-Relay, $\text{SR}^1 > 0$, when $h_{SE} < \min\{h_{SR}, \alpha h_{RD}\}$, and $h_{RE} < \min\{h_{RD}, \frac{h_{SR}}{\alpha}\}$.

C02: For FDJ, $\text{SR}^3 > 0$, when $h_{SE} < \min\{h_{SR} h_{RE}/h_{SI}, \alpha^2 h_{RD} h_{RE}/\sigma\}$, and $h_{RE} < \min\{h_{RD} h_{SE}/\sigma, h_{SR} h_{SE}/\alpha^2 h_{SI}\}$.

From C01 and C02, when $h_{SI} < h_{RE}$ or $h_{SI} < h_{SE}$, FDJ increases the chance of non-zero secrecy rate compared to that for HD-Relay protocol.

*Non-Zero Secrecy Rate Conditions for an Informed Eavesdropper:* For an informed eavesdropper, by considering $\gamma_E^2 < \gamma_{HD}$ and $\gamma_E^4 < \gamma_{FDJ}$, the conditions to have $SR^2 > 0$ and $SR^4 > 0$ are:

C03: $SR^2 > 0$ if $h_{SE} + \alpha h_{RE} < \min\{h_{SR}, \alpha h_{RD}\}$,
C04: $SR^4 > 0$ if $\frac{h_{SE}}{\alpha h_{RE}} + \frac{\alpha h_{RE}}{h_{SE}} < \min\{\frac{h_{SR}}{\alpha h_{SI}}, \frac{\alpha h_{RD}}{\sigma}\}$.

It is obvious that when $h_{SI} \leq \min\{h_{RE}, h_{SE}/\alpha\}$, the chance of non-zero secrecy rate is increased for FDJ.

**Proposition 1:** For a naive eavesdropper, FDJ achieves higher secrecy rate than HD-Relay in the following cases

C11: $h_{SR} < \alpha h_{RD} + \alpha^2 h_{RD} h_{SI}/\sigma$, $h_{SE} > \alpha h_{RE}$, and $h_{SI} < h_{RE}$,
C12: $h_{SR} < \alpha h_{RD} + \alpha^2 h_{RD} h_{SI}/\sigma$, $h_{SE} \leq \alpha h_{RE}$, and $\alpha h_{SI} < h_{SE}$,
C13: $h_{SR} \geq \alpha h_{RD} + \alpha^2 h_{RD} h_{SI}/\sigma$, and $\sigma < h_{SE}$.

*Proof:* See Appendix A. ∎

**Proposition 2:** For an informed eavesdropper, FDJ achieves higher secrecy rate than HD-Relay in the following cases

C21: $h_{SR} < \alpha h_{RD}$ and $h_{SI} \leq h_{RE} \times \frac{\beta^2 + \alpha\beta}{\beta^2 + \alpha^2}$,
C22: $h_{SR} \geq \alpha h_{RD}$ and $\frac{\sigma}{h_{RE}} \leq \frac{\alpha\beta(\alpha+\beta)}{(\alpha^2+\beta^2)}$,

where $\beta = \frac{h_{SE}}{h_{RE}}$.

*Proof:* See Appendix B. ∎

From C12 and C21, by adjusting $\alpha$ and $\frac{\beta^2+\alpha\beta}{\beta^2+\alpha^2}$, the upper bound for $h_{SI}$ is relaxed for naive and informed eavesdropper, respectively. For example, if $\beta \approx 1$, i.e., $h_{SE} \approx h_{RE}$, and $\alpha < 1$, the bound of $h_{SI}$ in C21 is relaxed, which is the very appealing from practical scenario for large $h_{SI}$.

## IV. POWER ALLOCATION PROBLEMS

The power allocation problem in a *multi-carrier* scenario for HD- and FD- Relay protocols to maximize the secrecy rate $SR^i$ under the S and R transmit power constraints can be stated in a general form as follows

$$\mathcal{O}^i : \max_{\mathbf{P} \geq 0} \quad SR^i$$

$$\text{s.t.} \quad C31: \sum_{k=1}^{K} P_S^k \leq P_S^{\max}, \quad C32: \sum_{k=1}^{K} P_R^k \leq P_R^{\max},$$

where $i = 1, \cdots, 4$ and $\mathbf{P} = [P_S^1, \cdots, P_S^K, P_R^1, \cdots, P_R^K]$ is the vector of transmit power of S and R for $k \in \mathcal{K}$. Each of the above optimization problems is non-convex and encounter high computational complexity. For solving them efficiently, we apply the following iterative algorithm based on DC programming.

Consider the non-convex function $\theta(\mathbf{P}) = \sum_{k=1}^{K} f^k(\mathbf{P}) - g^k(\mathbf{P})$ where $f^k(\mathbf{P})$ and $g^k(\mathbf{P})$ are the convex functions and $\mathbf{P}(l)$ is the power allocation vector at iteration $l = 0, 1, 2, \cdots$. In this case, $g^k(\mathbf{P}(l))$ is replaced with its first order Taylor expansion, i.e., $\tilde{g}^k(\mathbf{P}(l)) \approx g^k(\mathbf{P}(l-1)) + \nabla_{\mathbf{P}(l-1)} g^k(\mathbf{P}(l-1)) [\mathbf{P}(l) - \mathbf{P}(l-1)]^T$, where $\nabla_{\mathbf{P}(l-1)} g^k(\mathbf{P}(l-1))$ is the gradient vector of $g^k(\mathbf{P}(l-1))$ with respect to vector $\mathbf{P}(l-1)$. Now $\theta(\mathbf{P})$ can be approximated as

$$\theta(\mathbf{P}(l)) \approx \sum_{k=1}^{K} f^k(\mathbf{P}(l)) - \tilde{g}^k(\mathbf{P}(l)). \quad (1)$$

TABLE I
DC-BASED ITERATIVE ALGORITHM.

**Initialization**
Set $0 < \varepsilon \ll 1$, $l = 0$ and $\mathbf{P}(l = 0)$ as any feasible power vector;
**Iterations**: For a given $\mathbf{P}(l)$, execute three steps below:
 **Step 1**: Compute convex approximation from (1);
 **Step 2**: Solve the convex approximation and derive an optimal solution $\mathbf{P}(l)$;
 **Step 3**: If $\|\mathbf{P}(l) - \mathbf{P}(l-1)\| \leq \varepsilon$, stop. Otherwise, $l = l+1$ and go back to Step 1.

Now the right-hand side of (1) is the convex function since $\tilde{g}^k(\mathbf{P}(l))$ is linear function of $\mathbf{P}(l)$. The general DC-based iterative algorithm is given in Table I. In the following, we show that how this iterative algorithm can be applied to each of our specific optimization problems.

We first consider FDJ protocol for a naive eavesdropper with the following optimization problem $\mathcal{O}^3$

$$\max_{\mathbf{P} \geq 0} \sum_{k=1}^{K} \Big( \log 2(1 + \min\{\gamma_R^k(t=1), \gamma_D^k(t=2)\})) \quad (2)$$
$$- \log 2(1 + \max\{\gamma_{SEJ}^k(t=1), \gamma_{REJ}^k(t=2)\}) \Big)$$
s.t. C31 and C32.

We rewrite (2) as

$$\max_{\boldsymbol{\pi}>0, \boldsymbol{\varpi}>0, \mathbf{P} \geq 0} \sum_{k=1}^{K} (\pi^k - \varpi^k) \quad (3)$$

s.t. C31, and C32,
C33: $\log 2(1 + \gamma_R^k(t=1)) \geq \pi^k$,
C34: $\log 2(1 + \gamma_D^k(t=2)) \geq \pi^k$
C35: $\log 2(1 + \gamma_{SEJ}^k(t=1)) \leq \varpi^k$,
C36: $\log 2(1 + \gamma_{REJ}^k(t=2)) \leq \varpi^k$,

where $\boldsymbol{\pi} = [\pi^1, \cdots, \pi^K]$ and $\boldsymbol{\varpi} = [\varpi^1, \cdots, \varpi^K]$. To use the algorithm in Table I, the convexified version of (3) for iteration $l$ is

$$\max_{\boldsymbol{\pi}>0, \boldsymbol{\varpi}>0, \mathbf{P}(l) \geq 0} \sum_{k=1}^{K} (\pi^k - \varpi^k) \quad (4)$$

s.t. C31, and C32,
C43: $\pi^k \leq f_R^k(\mathbf{P}(1,l)) - g_R^k(\mathbf{P}(1,l-1)) - \frac{\partial g_R^k(\mathbf{P}(1,l-1))}{\partial P_R^k}(P_R^k(1,l) - P_R^k(1,l-1))$,
C44: $\pi^k \leq f_D^k(\mathbf{P}(2,l))$,
C45: $f_{SE}^k(\mathbf{P}(1,l)) - g_{SE}^k(\mathbf{P}(1,l-1)) - \frac{\partial g_{SE}^k(\mathbf{P}(1,l-1))}{\partial P_R^k}(P_R^k(1,l) - P_R^k(1,l-1)) \leq \varpi^k$,
C46: $f_{RE}^k(\mathbf{P}(2,l)) - g_{RE}^k(\mathbf{P}(2,l-1)) - \frac{\partial g_{RE}^k(\mathbf{P}(2,l-1))}{\partial P_S^k}(P_S^k(2,l) - P_S^k(2,l-1)) \leq \varpi^k$,

where the functions related to C43-C46 are

$$f_R^k(\mathbf{P}(1,l)) = \log 2(\sigma + P_R^k(1,l)h_{SI}^k + P_S^k(1,l)h_{SR}^k),$$
$$f_{SE}^k(\mathbf{P}(1,l)) = \log 2(\sigma + P_R^k(1,l)h_{RE}^k + P_S^k(1,l)h_{SE}^k),$$
$$f_{RE}^k(\mathbf{P}(2,l)) = \log 2(\sigma + P_R^k(2,l)h_{RE}^k + P_S^k(2,l)h_{SE}^k),$$
$$f_D^k(\mathbf{P}(2,l)) = \log 2(1 + \frac{P_R^k(2,l)h_{RD}^k}{\sigma}),$$
$$g_R^k(\mathbf{P}(1,l-1)) = \log 2(\sigma + P_R^k(1,l-1)h_{SI}^k),$$
$$g_{SE}^k(\mathbf{P}(1,l-1)) = \log 2(\sigma + P_R^k(1,l-1)h_{RE}^k),$$
$$g_{RE}^k(\mathbf{P}(2,l-1)) = \log 2(\sigma + P_S^k(2,l-1)h_{SE}^k).$$

where $P_X^k(1,l)$ and $P_X^k(1,2)$ are the transmit power of X = {S,R} at $t=1$ and $t=2$ for sub-carrier $k$ at iteration $l$, respectively; and we have $\mathbf{P}(1,l) = \{P_X^1(1,l), \cdots P_X^K(1,l)\}$ and $\mathbf{P}(1,2) = \{P_X^1(1,2), \cdots P_X^K(1,2)\}$. The optimization problem $\mathcal{O}^4$, related to FDJ for an informed eavesdropper, is

$$\max_{\boldsymbol{\pi} > 0, \boldsymbol{\varpi} > 0, \mathbf{P}(l) \geq 0} \sum_{k=1}^{K} (\pi^k - \varpi^k), \quad (5)$$
$$\text{s.t.:} \quad \text{C31, C32, C33, C34,}$$
$$\text{C38: } \log 2(1 + \gamma_E^{k,6}) \leq \varpi^k.$$

Here, again, we can utilize C43 and C44 to transform (5) into its convex approximation. Therefore, the DC-approximation of above problem for iteration $l$ is

$$\max_{\boldsymbol{\pi} \geq 0, \boldsymbol{\varpi} \geq 0, \mathbf{P}(l) \geq 0} \sum_{k=1}^{K} (\pi^k - \varpi^k), \quad (6)$$
$$\text{s.t.:} \quad \text{C31, C32, C43, C44 and C48,}$$

where C48 is $f_E^{k,4}(\mathbf{P}(l)) - g_E^{k,4}(\mathbf{P}(l-1)) - \frac{\partial g_E^{k,4}(\mathbf{P}(l-1))}{\partial P_R^k(1,l-1)}(P_R^k(1,l) - P_R^k(1,1-1)) - \frac{\partial g_E^{k,4}(\mathbf{P}(l-1))}{\partial P_S^k(2,l-1)}(P_S^k(2,l) - P_S^k(2,1-1)) \leq \varpi^k$, where $f_E^{k,4}(\mathbf{P}(l)) = \log 2\big((\sigma + P_R^k(1,l)h_{RE}^k)(\sigma + P_S^k(2,l)h_{SE}^k) + (P_S^k(2,l)h_{SE}^k)(\sigma + P_S^k(2,l)h_{SE}^k) + (P_R^k(1,l)h_{RE}^k)(\sigma + P_R^k(1,l)h_{RE}^k)\big)$ and $g_E^{k,4}(\mathbf{P}(l-1)) = \log 2((\sigma + P_R^k(1,l-1)h_{RE}^k)(\sigma + P_S^k(2,l-1)h_{SE}^k))$.

The formulations related to $\mathcal{O}^3$ and $\mathcal{O}^4$ can be used for the optimization problems of the HD-Relay protocol for the naive and informed eavesdropper, respectively, except that the transmit power of R in $t=1$ and the transmit power of S in $t=2$ must be set to zero.

## V. ILLUSTRATIVE EXAMPLES

To investigate the performances of FDJ, we conduct the following simulations. In simulation setup, the channel gains are assumed to be i.i.d. with Rayleigh-distributed, i.e., $h_{XY}^k = \mathcal{CN}(0, 1/d_{XY}^\varsigma)$ for each $k \in \mathcal{K}$ where $d_{XY}$ is the distance between X and Y, $\varsigma = 4$ is the path-loss exponent. The location of S, R and D are on $(0,0)$, $(1,0)$ and $(2,0)$, and E is located between S and R and $d_{SE} = d_{RE}$. The SR-to-SE distance ratio, $d_{SR}/d_{SE}$, is used as a reference parameter in presenting the simulation results. We set $K = 16$, $\varepsilon = 0.001$ for the algorithm in Table I, and $P_S^{\max} = P_R^{\max} = 5$ unless otherwise stated. We compare the performance of FDJ and HD-Relay. The secrecy rate for each scheme is measured during each transmission frame, and the results are derived from the average over 1000 channel realizations.

In Figs. 2(a) and 2(b), $SR^1$, $SR^2$, $SR^3$ and $SR^4$ versus $\frac{d_{SR}}{d_{SE}}$ and $\rho$ are shown, respectively. We see that with increasing $\frac{d_{SR}}{d_{SE}}$, decreasing $\frac{h_{SR}}{h_{SE}}$, the secrecy rate is decreased. From Fig. 2(a), when $\frac{d_{SR}}{d_{SE}} \leq 0.5$ and $\rho \geq 0.3$, e.g., C11 and C12 in Proposition 1 do not hold. Consequently, the FDJ secrecy rate is lower than those of HD-Relay.

As shown in Figs. 2(a) and 2(b), $SR^3$ and $SR^4$ are much less sensitive to $\frac{d_{SR}}{d_{SE}}$ than the secrecy rate for HD-Relay. This is because, FDJ can adjust its transmit power at $t=1$ and $t=2$ such that the secrecy rate is maintained even for large $h_{SE}$, i.e., $\frac{d_{SR}}{d_{SE}} = 1.1$.

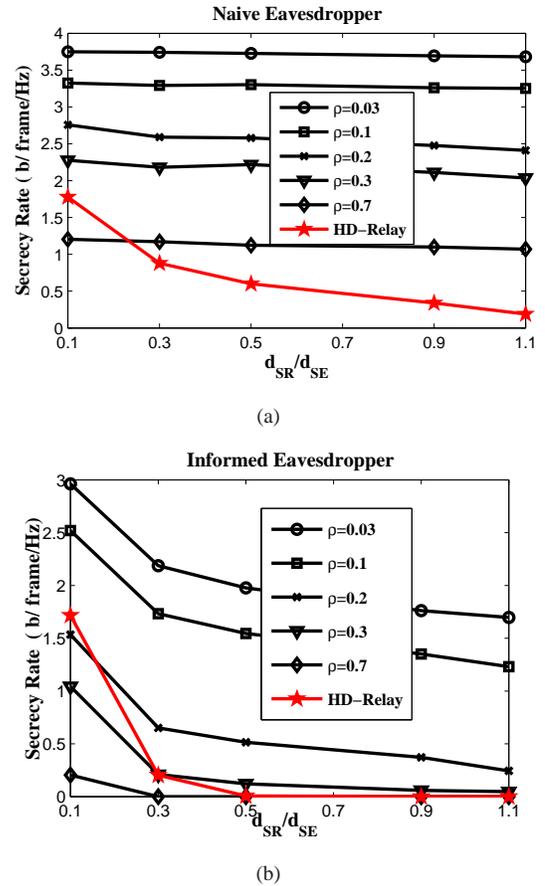

Fig. 2. FDJ and HD-Relay secrecy rates versus $\frac{d_{SR}}{d_{SE}}$ for various values of $\rho$ and (a) naive eavesdropper and (b) informed eavesdropper.

Comparing Fig. 2(a) and Fig. 2(b) shows that $SR^4$ is much more sensitive to $\rho$ than $SR^3$, as we expected from Propositions 1-2, i.e., the upper bound of $h_{SI}$ in C11 and C12 are more relaxed compared to the upper bound of $h_{SI}$ in C21. Also, Fig. 2(a) shows that for $\rho \geq 0.3$, HD-Relay achieve a better secrecy rate than FDJ, which is in line with C21 in Proposition 2, i.e., when $h_{SI} \geq h_{RE} \times \frac{\beta^2 + \alpha\beta}{\beta^2 + \alpha^2}$, $SR^4 < SR^2$.

Figs. 3(a)-3(b) depict the effects of increasing $P_S^{\max}$ on FDJ secrecy rate via parameters $\eta_S = \frac{SR^3}{[SR^3]_{P_S^{\max}=5}}$ as a normalized

secrecy rate. As shown in Fig. 3(a), $SR^3$ is not interference-limited, which is a desirable feature of FDJ. However, shown in Fig. 3(b), $P_S^{\max}$ does not have considerable effect on the decreasing or increasing $SR^4$.

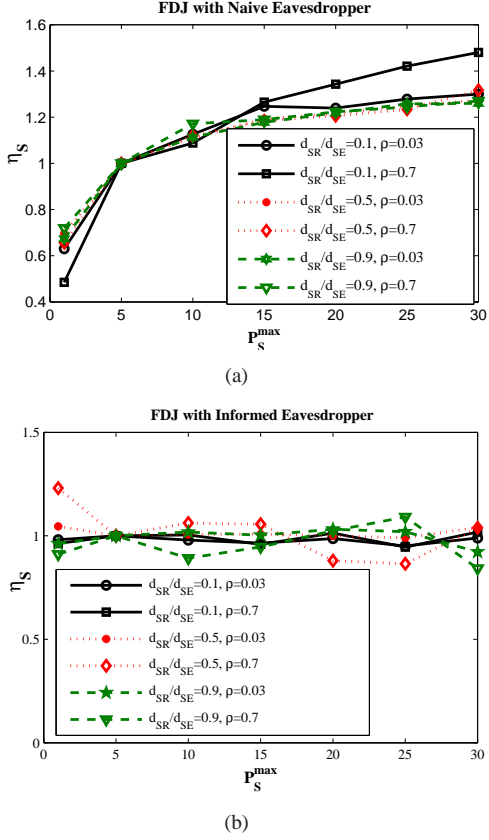

Fig. 3. FDJ $\eta_S$ versus $P_S^{\max}$ for (a) naive eavesdropper and (b) informed eavesdropper.

## VI. CONCLUSION

We studied the secure transmission between the source, the relay with full-duplex (FD) capability, and the destination in the presence of the informed or naive eavesdroppers. We proposed FD with jamming (FDJ) protocol where first, FD-Relay receives the data and sends the jamming to eavesdropper, and, then, the FD-Relay sends the data to destination while source jams the eavesdropper. Via the analysis and simulation results, the achieved secrecy rates of the proposed FDJ and conventional half-duplex (HD)-Relay protocols are compared. In summary, the achieved secrecy rate of FDJ can be up to twice that of HD-Relay for small level of self interference (SI) of FD transmission.

s

### A. Proof of Proposition 1

From assumptions in Section III, $\gamma_R(t=1) = h_{SR}/\alpha h_{SI}$, $\gamma_D(t=2) = \alpha h_{RD}/\sigma$, $\gamma_{SEJ}(t=1) = h_{SE}/\alpha h_{RE}$, $\gamma_{REJ}(t=2) = h_{RE}/\alpha h_{SE}$, $\gamma_R(HD) = h_{SR}/\sigma$, $\gamma_D(HD) = \alpha h_{RD}/\sigma$, $\gamma_{SE}(HD) = h_{SE}/\sigma$, and $\gamma_{RE}(HD) = h_{RE}/\sigma$. We want to derive the conditions for $SR^3 > SR^1$. When $h_{SR} < \alpha h_{RD} + \alpha^2 h_{RD} h_{SI}/\sigma$, $\gamma_{FDJ} = \min\{\gamma_R(t=1), \gamma_D(t=2)\} = \gamma_R(t=1) = h_{SR}/\alpha h_{SI}$. If $h_{SE} > \alpha h_{RE}$, $\gamma_E^3 = \max\{\gamma_{SEJ}(t=1), \gamma_{REJ}(t=2)\} = \gamma_{SEJ}(t=1) = h_{SE}/(\alpha h_{RE})$, $SR^3 = \log 2(\frac{h_{SR} h_{RE}}{h_{SI} h_{SE}})$ and $SR^1 = \log 2(\frac{h_{SR}}{h_{SE}})$. When $h_{SI} < h_{RE}$, $\log 2(\frac{h_{SR} h_{RE}}{h_{SI} h_{SE}}) > \log 2(\frac{h_{SR}}{h_{SE}})$, i.e., C11. If $h_{SE} \le \alpha h_{RE}$, $\gamma_E^3 = \max\{\gamma_{SEJ}(t=1), \gamma_{REJ}(t=2)\} = \gamma_{REJ}(t=1) = \log 2(\alpha h_{RE}/h_{SE})$. Therefore, $SR^3 = \log 2(\frac{h_{SR} h_{SE}}{\alpha^2 h_{SI} h_{RE}})$ and $SR^1 = \log 2(\frac{h_{SR}}{\alpha h_{RE}})$. To have $\log 2(\frac{h_{SR} h_{SE}}{\alpha^2 h_{SI} h_{RE}}) > \log 2(\frac{h_{SR}}{\alpha h_{RE}})$, we should have $h_{SI} < h_{SE}/\alpha$, i.e., C12. When $h_{SR} < \alpha h_{RD} + \alpha^2 h_{RD} h_{SI}/\sigma$, $\gamma_{FDJ} = \min\{\gamma_R(t=1), \gamma_D(t=2)\} = \gamma_D(t=2) = \alpha h_{RD}/\sigma$. If $h_{SE} \le \alpha h_{RE}$, $SR^3 = \log 2(\frac{h_{RD}\alpha}{\sigma} \times \frac{\alpha h_{RE}}{h_{SE}})$ and for the HD-Relay, $SR^1 = \log 2(\frac{h_{RD}\alpha}{\sigma} \times \frac{\sigma}{h_{SE}})$. Now, for $SR^3 > SR^1$, we must have $\alpha h_{RE} > \sigma$. From $h_{SE} \le \alpha h_{RE}$, the latter inequality is transformed into $\sigma < h_{SE}$. If $h_{SE} > \alpha h_{RE}$, $SR^3 = \log 2(\frac{h_{RD}\alpha}{\sigma} \times \frac{\alpha h_{RE}}{h_{SE}})$ and $SR^1 = \log 2(\frac{h_{RD}\alpha}{\sigma} \times \frac{\sigma}{h_{SE}})$. If $\sigma < \alpha h_{RE}$, $SR^3 > SR^1$. Since $h_{SE} > \alpha h_{RE}$, we should have $\sigma < h_{SE}$. Therefore, C13 is derived.

### B. Proof of Proposition 2

We want to derive the conditions for $SR^4 > SR^2$. Now, $\gamma_E^4 = \frac{h_{SE}}{\alpha h_{RE}} + \frac{\alpha h_{RE}}{h_{SE}} = \frac{\beta}{\alpha} + \frac{\alpha}{\beta}$. If $h_{SR} < \alpha h_{RD}$, $SR^4 = \log 2(\frac{h_{SD}}{\alpha h_{SI} \gamma_E^4})$ and $SR^2 = \log 2(\frac{\alpha h_{RD}}{\sigma} \times \frac{\sigma}{\alpha h_{RE} + h_{SE}})$. For $SR^4 > SR^2$, we need $\log 2(\frac{h_{SD}}{\alpha h_{SI} \gamma_E^4}) > \log 2(\frac{\alpha h_{RD}}{\sigma} \times \frac{\sigma}{\alpha h_{RE} + h_{SE}})$. With some rearrangements, C21 is derived. If $h_{SR} \ge \alpha h_{RD}$, $SR^4 = \log 2(\frac{\alpha h_{RD}}{\sigma \gamma_E^4})$ and $SR^1 = \log 2(\frac{\alpha h_{RD}}{\sigma} \times \frac{\sigma}{\alpha h_{RE} + h_{SE}})$. For $SR^4 > SR^1$, we need $\log 2(\frac{h_{SD}}{\alpha h_{SI} \gamma_E^4}) > \log 2(\frac{\alpha h_{RD}}{\sigma} \times \frac{\sigma}{\alpha h_{RE} + h_{SE}})$. With some rearrangements, C22 is derived.